# Fabrication and cold test of prototype of spatially periodic radio frequency quadrupole focusing linac

Peiyan Yu[1,2], Bin Zhang[1,*], Fengfeng Wang[1], Chenxing Li[1], Guozhen Sun[1],
Zhijun Wang[1], Lubei Liu[1,2], Chenzhang Yuan[1], Yuan He[1], Hushan Xu[1]

**Affiliations**

[1]Institute of Modern Physics, Chinese Academy of Sciences, Lanzhou 730000, China

[2]School of Nuclear Science and Technology, University of Chinese Academy of Sciences, Beijing 100049, China

[*]Corresponding author: zhangb@impcas.ac.cn (Bin Zhang)

**Abstract**

    A 325 MHz aluminum prototype of a spatially periodic rf quadrupole focusing linac was developed at the Institute of Modern Physics, Chinese Academy of Sciences, as a promising candidate for the front end of a high-current linac. It consists of an alternating series of crossbar H-type drift tubes and rf quadrupole sections. Owing to its special geometry, cavity fabrication is a major hurdle for its engineering development and application. In this paper, we report the detailed mechanical design of this structure and describe its fabrication process, including machining, assembly, and inspection. The field distribution was measured by the bead-pull technique. The results show that the field errors of both the accelerating and focusing fields are within an acceptable range. A tuning scheme for this new structure is proposed and verified. The cold test process and results are presented in detail. The development of this prototype provides valuable guidance for the application of the spatially periodic rf quadrupole structure.

**Body of the paper:**

## 1. Introduction

    Accelerating structures with electric field focusing are routinely used in the front end of high-current proton or heavy-ion linacs. For example, the spatially uniform rf quadrupole (RFQ) focusing structure has become the optimal choice for use after the ion source because it integrates beam acceleration, bunching, and transverse focusing in one compact cavity [1,2]. However, the accelerating efficiency of the RFQ structure decreases as the beam energy increases because the longitudinal electric field is provided only by vane modulation. Therefore, the RFQ structure is generally used for beam energies ranging from tens of keV to several MeV.

    To extend the application range of the RFQ focusing principle, several more efficient accelerating structures have been developed as alternatives to the higher-energy section of an RFQ linac and its subsequent accelerating structures; these structures are obtained by combining the RFQs with accelerating gaps formed between drift tubes. An early structure proposed by V.A. Teplyakov [3]

supplements the accelerating gap with four conducting fingers arranged in a quadrupole geometry, which generate a focusing field in the accelerating gap. Structures based on this principle have been employed for many years in a linac injector for the proton synchrotron at the Institute for High Energy Physics [4]. A comparable scheme is the rf-focused drift tube (RFD) linac structure, in which the RFQ focusing field is placed inside a drift tube by splitting the drift tube into two independent electrodes that support two fingers each [5]. As a modified version of the RFD linac, the rf-focused interdigital linac structure adopts a combination of the interdigital H-mode structure and the RFQ focusing structure to reduce the cavity size and improve the shunt impedance [6].

In recent years, new structures with spatially periodic RFQ focusing have been developed, which are characterized by the presence of separate accelerating and focusing zones. This concept was first suggested in the Hybrid RFQ for the acceleration of low-charge-state heavy-ion beams [7,8]. In addition, a structure suitable for all ion species was proposed; it consists of an alternating series of crossbar H-type drift tubes and four-vane-type RFQ sections. A detailed beam dynamics study showed that this structure can provide a high energy gain rate and high shunt impedance with moderate transverse emittance growth [9]. The electrode geometry is somewhat simpler than the focusing finger geometry; it avoids an overly high surface electric field and reduces the difficulty of rf tuning. In particular, this structure can be built using the conventional four-vane-type RFQ cavity, which makes it easier to cool and affords the potential for continuous-wave operation.

The spatially periodic RFQ structure is a promising choice for the design of a compact and efficient front end of a high-current linac. Consequently, the research and development of this structure was begun at the Institute of Modern Physics, Chinese Academy of Sciences, as a pre-study of the High Intensity heavy ion Accelerator Facility (HIAF) project. The HIAF is being designed to provide intense primary and radioactive ion beams for nuclear physics, atomic physics, and application research [10]. The Ion-Linac used as the injector in the HIAF is a superconducting (SC) heavy-ion accelerator; it consists of a low-energy beam transport line, an RFQ, a medium-energy beam transport line, and an SC section [11,12]. The spatially periodic RFQ structure is intended to serve as an intermediate structure between the RFQ and SC cavities. It can be used with a shorter RFQ and improve the input energy of the SC section. As a key step in the study, an aluminum prototype was designed and fabricated to investigate the feasibility of the mechanical design and process technology for the new structure and also to determine the field distribution and its tunability. In this paper, the mechanical design, rf measurement, and tuning of the prototype are presented in detail.

## 2. RF structure

The rf structure of the prototype cavity is based on that of a conventional four-vane-type RFQ in which the tips of the vanes are cut off, and the remaining parts make up four girders. Drift tubes connected to two stems are installed crosswise on the girders to form the drift tube linac (DTL) sections. The RFQ sections are built by mounting the electrode blocks on the girders at the designed positions. Figure 1a shows the configuration of the alternating DTL and RFQ sections of the prototype cavity.

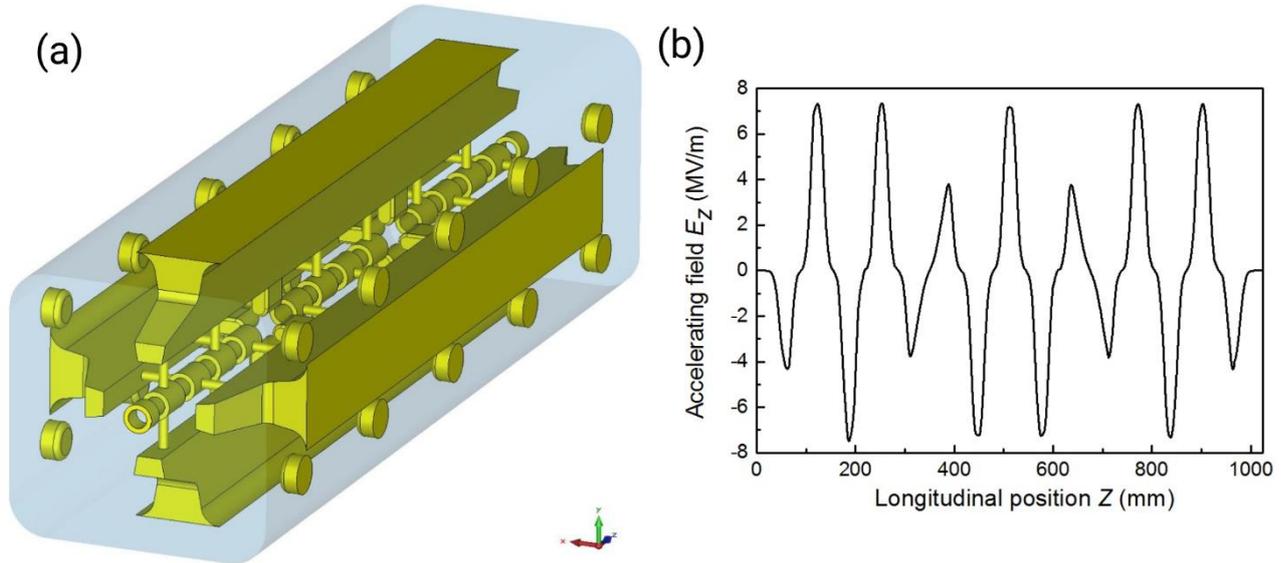

Fig. 1 **a** RF structure of the prototype, **b** distribution of the accelerating field along the central axis of the cavity with 1 J of stored energy

The cavity is operated in an H210-like mode, and no dipole modes exist near the operating mode because the drift tubes cause the opposite girders to be electrically shorted [9,13]. The longitudinal electric field is plotted in Fig. 1b, which shows that charged-particle acceleration will take place in both the gaps between the conventional drift tubes and the gaps between the drift tubes and RFQs. The RFQs provide transverse focusing for the beam and can also provide acceleration if the electrodes are modulated. The main parameters of the RF model are summarized in Table 1. The frequency of the prototype cavity was set to 325 MHz to keep its size relatively small. The cavity has 14 drift tubes (including two in the endplates) and two unmodulated RFQs. Twenty slug tuners with a diameter of 40 mm were arranged in four quadrants to study the tunability of the cavity. The cells were set to be identical in length to obtain a naturally flat field distribution and to simplify fabrication.

Table 1 Main parameters of the rf model

| Parameter | Value |
| --- | --- |
| Frequency (MHz) | 325 |
| Cavity length (mm) | 1025 |
| Transverse dimension (mm) | $278.2 \times 278.2$ |
| Number of drift tubes | 5-4-5 |
| Drift tube aperture radius (mm) | 10 |
| RFQ aperture radius (mm) | 15 |
| $Q$ value of the aluminum cavity | 9289 |

## 3. Mechanical structure design and fabrication

### 3.1. Mechanical design

As shown in Fig. 2a, the main structure of the prototype cavity is a box resembling that employed for the RFQ of the C-ADS Injector Ⅱ [14]. The external structure of the cavity is divided into four cavity walls and two endplates. The four girders are integrated with the four corresponding cavity walls. Inside the cavity, the drift tubes and RFQs are distributed alternately on top of the girders. Each drift tube is connected with girders on either side by two collinear stems, and the stems of adjacent drift tubes are perpendicular to each other.

The cavity is an overconstrained structure owing to the configuration of the stems inside the cavity; that is, one or more degrees of freedom (DOFs) of the components are repeatedly constrained by different fixtures. In the mechanical design process, a drift tube connected to a stem at each end is referred to as a drift tube component to avoid confusion with individual drift tubes. To install one drift tube component, the stems on each end have to be inserted into the installation holes on each symmetrical girder, as shown on the left side of Fig. 2b. The DOFs of the drift tube component, except for the translational DOF in the axial direction of the stem, are overconstrained by the two installation holes. In turn, each cavity wall is subject to redundant constraints from the drift tube components, as it is fully constrained by the other cavity walls and endplates. The RFQ section is formed by inserting electrode blocks into the installation holes of the girders. Consequently, the single-electrode block is underconstrained because the translational DOF in the installation direction is free. The fabrication of the overconstrained structure requires high-precision machining because insufficient machining precision tends to cause assembly failure.

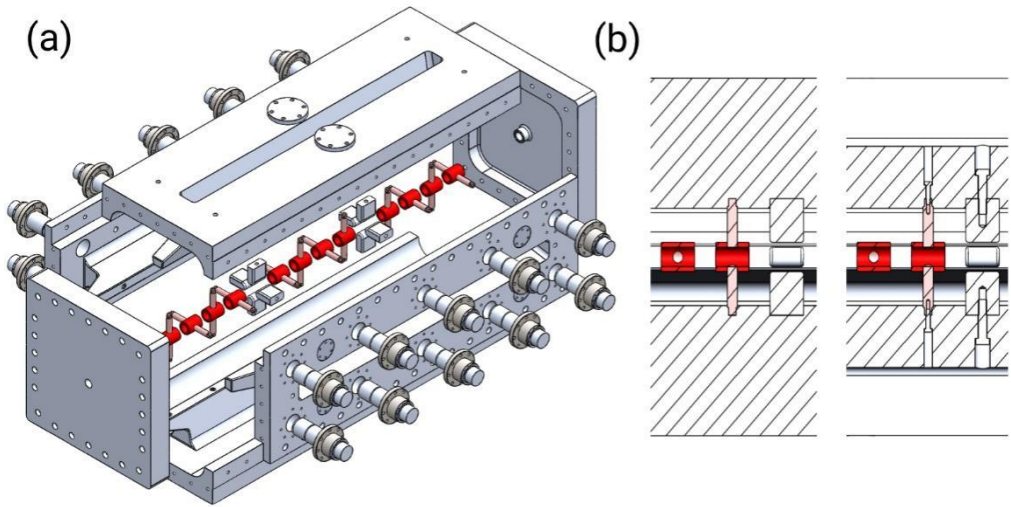

Fig. 2 **a** Mechanical model of the prototype cavity, **b** illustration of the positioning of the drift tube component without threaded connection (left) and with threaded connection (right)

One of the major differences between the proposed prototype cavity and the conventional DTL cavity is that the stems are positioned on the girders instead of on the outer cavity walls, which means that the drift tube component cannot be operated manually in certain assembly steps. During the assembly of the third or fourth cavity wall, the six installation holes on the girder must be fitted with the six drift tube components mounted on the opposite girder. Even if all the critical dimensions are guaranteed to be within tolerance, there is still a high risk of assembly failure without manual operation.

Owing to the mechanical structure of the prototype cavity, threaded connections are added to the drift tube components and the electrode blocks, as shown on the right side of Fig. 2b. A groove is designed on the cavity wall to enable the use of short, stable bolts. The remaining free DOF of the electrode block is constrained by the bolt. For the drift tube components, the constraint on the translational DOF in the axial direction of the stem is transferred from the bottom of the installation holes to the groove on the cavity wall. In addition, the overconstraint on the cavity wall in this direction is removed. Furthermore, this design makes it possible to manually control the drift tube components during the assembly process, thereby reducing the risk of assembly failure.

### 3.2. Fabrication

In the processing of the prototype cavity, several methods are used to control the machining precision of the parts. The concentricity of the stems in the drift tube component is a crucial parameter and strongly affects the overall assembly precision of the cavity. Therefore, a unique processing method is proposed to fabricate the drift tube components, as shown in Fig. 3a. First, the external cylindrical surface of the drift tube is machined to its design size, and a radial hole for the connection with the stem and an axial hole as the center reference for subsequent processing are drilled. Next, the stem is penetrated through the drift tube using cold assembly technology. Finally, the internal cylindrical surface of the drift tube is machined to its design size; during this process, the long stem is split into two parts. The magnitude of the interference is set to 0.08 mm to ensure that the connection between the drift tube and the stem is reliable. Coordinate measuring machine results show that the concentricity of the stems in the drift tube component can reach $\phi 0.05$ mm.

The machining of the aluminum alloy is likely to generate intense local stress and deformation, which would cause the workpiece to be scrapped for being out of tolerance. To reduce the machining deformation, the milling of the cavity walls and endplates proceeded as follows: rough machining, semi-finishing, and finishing. To eliminate the residual stress, heat treatment was applied after rough machining, and natural aging was performed after the other process steps.

### 3.3. Assembly and inspection

The assembly of the prototype cavity was performed under the guidance of the in-process inspection with a coordinate measuring arm [15]. Before the assembly process, the mating dimensions of the four cavity walls were inspected individually. The results showed that only the length of the cavity wall is out of tolerance owing to thermal shrinkage, which is relatively large (-0.3 to -0.4 mm) but would not hinder the assembly process. Then the drift tube components were mounted on the corresponding girder, the concentricity and spacing dimensions of which were measured. The measurement results can provide a preliminary check on the fitting accuracy and verify the feasibility of assembly.

In the assembly process, the bottom and left cavity walls, each of which was equipped with six drift tube components, were initially assembled, and the right cavity wall was installed (see Fig. 3b). When the girder of the right cavity wall approached the drift tube components mounted on the opposite girder, a thread rod was used to pull the drift tube components out to align with both girders. After all six components were mounted on the two girders, the right cavity wall was slowly pushed into the design position, and the three assembled cavity walls formed a U-type subassembly. The top cavity wall was assembled by the same process. Fig. 3c shows the assembled cavity without the front endplate.

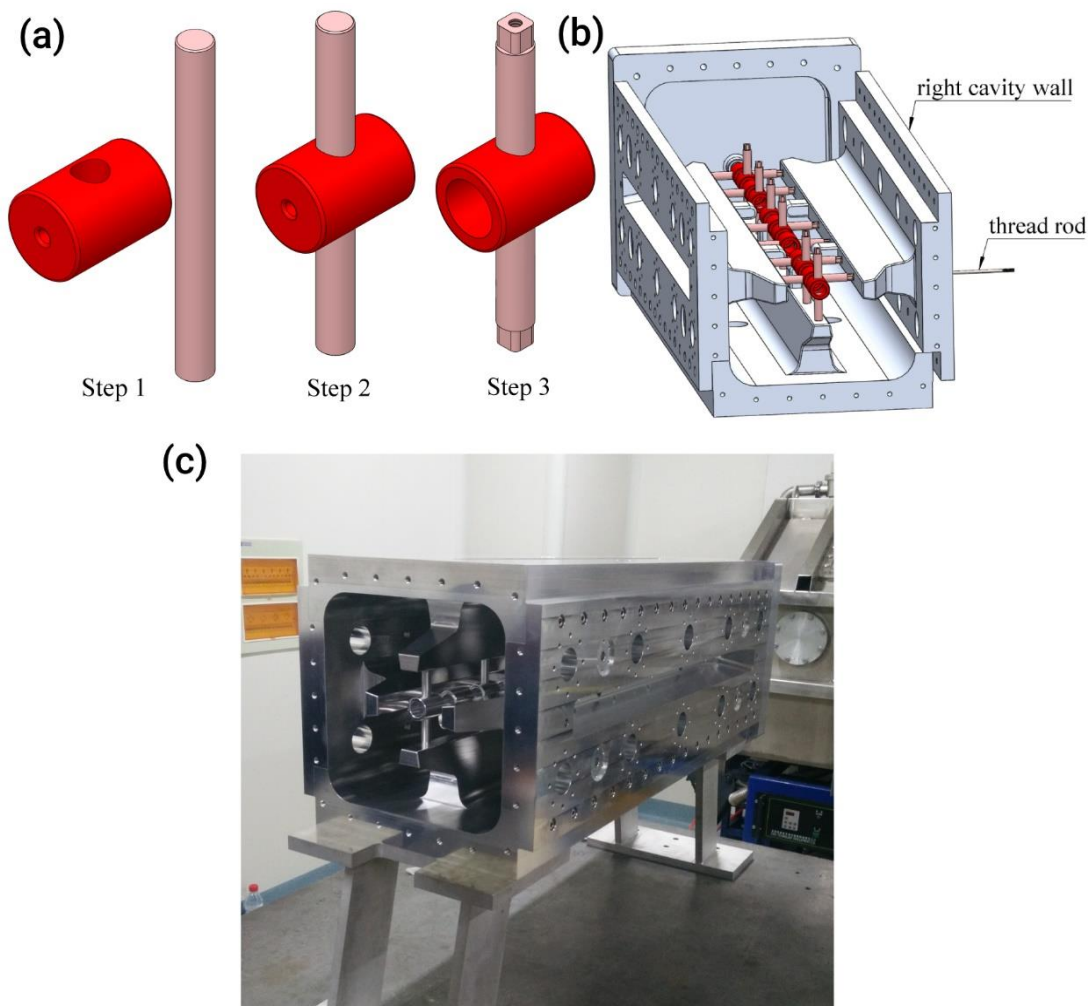

Fig. 3 Fabrication and assembly process: **a** processing of drift tube component, **b** assembly process of right cavity wall, **c** assembled cavity without the front endplate

The fabrication precision of the cavity determines the concentricity of the drift tubes, which is a critical parameter for efficient beam transmission and is hard to measure for the finished cavity, because the probe of the coordinate measuring arm cannot reach the drift tubes in the middle of the cavity. Therefore, the concentricity of the drift tubes was initially measured on the U-type subassembly, and the results are shown in Fig. 4. Note that the measured results differ slightly from those of the complete assembly owing to the absence of the top cavity wall. In the U-type subassembly, the measured positions of the vertical drift tube components (drift tubes 1, 3, 6, 8, 9, and 11) are slightly lower than the design positions because these components are bolted at the bottom and unconstrained at the top. Conversely, the positions of the horizontal drift tube components (drift tubes 2, 4, 5, 7, 10, and 12) are slightly higher than the design positions because the left and right cavity walls are not precisely parallel but have a small open angle under the constrained condition.

A scheme for measuring the concentricity of the drift tubes in the finished cavity using a laser tracker was proposed. A rod with a magnet attached to the front end was designed that can fix a ferromagnetic spherically mounted retroreflector (SMR) by the attractive magnetic force. By using this rod, the SMR could reach the surfaces of all the drift tubes to perform the measurement with the laser tracker. The measurement results are shown in Fig. 4. Taking the central axis defined by the cavity's internal surfaces as a reference, the concentricity of the twelve drift tubes is $\phi 0.22$ mm; that is, the maximum deviation from the central axis is 0.11 mm. This result will be taken into account in the error study of the beam dynamics design, and the technical process will be optimized further to improve the precision.

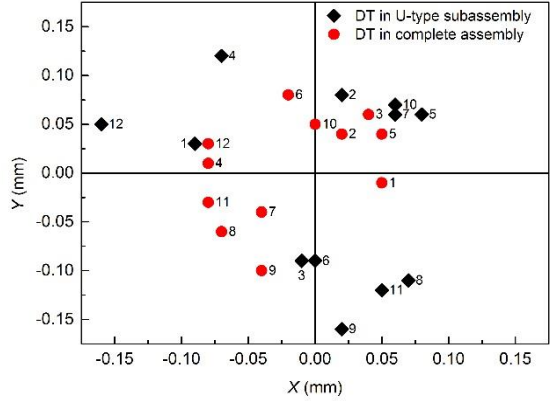

Fig. 4 Measurement results of the drift tube (DT) positions using the central axis as a reference

## 4. RF measurement

### 4.1. Experimental setup

Although much can be learned from the cavity field calculation, it is still necessary to measure a prototype cavity to determine the field distribution, especially for a new structure with a relatively complex field pattern. The bead-pull technique is used for the rf measurement. According to the Slater perturbation theorem, the measured electric field amplitude is proportional to the square root of the frequency shift induced by a dielectric sphere, $E \propto \sqrt{\Delta \omega_0}$ [16,17]. A photograph of the measurement apparatus is shown in Fig. 5. A bead tied to a string is driven by a step motor to move through the cavity, and the cavity frequency, as monitored by a vector network analyzer (VNA), is collected simultaneously with the step number of the motor.

In our early measurements, we found many errors caused by the noise floor of the VNA at the bottom of the field distribution curve. This phenomenon can be explained as follows. First, the relatively low $Q$ value of the aluminum cavity results in a wider bandwidth, and more noise is collected during the measurement. In addition, according to the relationship between the electric field amplitude and the frequency shift, the noise will give rise to a larger deviation in the low-field areas, for example, in the drift tubes [18].

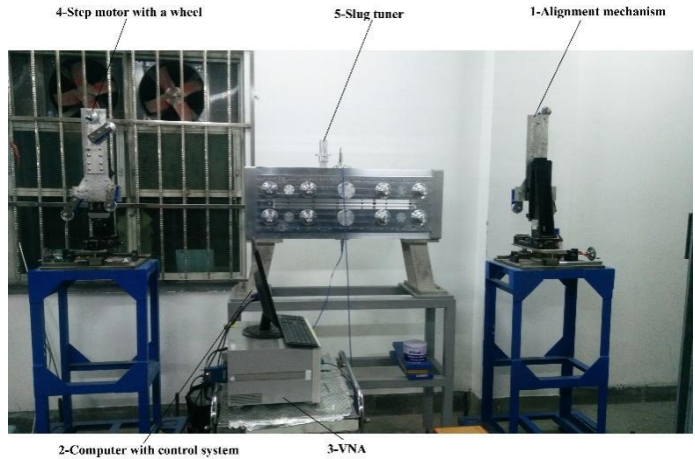

Fig. 5 Photograph of the bead-pull measurement apparatus

To improve the measurement accuracy, the noise floor of the VNA was reduced by combining the averaging and the intermediate-frequency bandwidth; these two noise reduction features of the VNA are commonly used to address the measurement issue mentioned above [19]. However, the use of both features prolongs the measurement time, during which environmental parameters such as temperature and humidity might change, introducing errors to the measurements [20]. Therefore, preliminary tests were performed to determine an appropriate setting of the VNA at which both the noise floor power and the measurement time are acceptable. Several dielectric spheres with different diameters were used as the perturbation bodies in these tests. It was found that the relative amplitude of the errors decreased with increasing sphere diameter, because a larger sphere would produce a larger frequency shift. However, this finding does not mean that the perturbation body should be as large as possible. The reason will be explained in Section 4.2.1. The perturbation body chosen for the field distribution measurement is a dielectric sphere with a diameter of 6 mm, and the maximum frequency shift it produces is approximately 7 kHz.

## 4.2. Results of RF measurement

The frequency, the $Q$ value, and the distribution of the accelerating and focusing fields of the prototype cavity were measured with all the tuners adjusted to the design insertion depth of 20 mm. The measured frequency was 326.72 MHz, which is 1.72 MHz higher than the design value. This deviation is attributed mainly to the errors in the frequency solution and fabrication errors. Thermal shrinkage makes the major contribution to the fabrication errors because the cavity was not processed in a constant temperature environment. The measured $Q$ value was 6061 when no rf seals were used, which is 65% of the simulated value.

### 4.2.1. Distribution of the accelerating field

Figure 6a compares the measured and simulated field distributions along the central axis of the cavity, which show good agreement overall. The field amplitudes in the gaps between the internal drift tubes are flat. The normalized amplitude reaches 0.6 in the gaps at both ends of the cavity, and it is approximately 0.5 in the gaps between the drift tubes and the RFQs. To quantify the uniformity of the overall accelerating field, the amplitude deviation in the nine gaps between the internal drift tubes is defined as the field unevenness, which is 1.3%.

In addition, inconsistencies appear between the measured and simulated results in the two RFQ sections. According to the simulated results, the amplitude should decrease to zero in the middle of the RFQ section, whereas the measured value at this position is 0.22. To explain this deviation, the simulated distribution of the transverse field at this position is investigated, as shown in Fig. 6b. The amplitude along the angle bisector of the quadrants is zero at the center, but it increases linearly with increasing distance from the center. The Slater perturbation theorem assumes that the unperturbed field at the measurement position is homogeneous [17]. In our experiment, the measured field amplitude corresponds to the average amplitude over the 6 mm sphere, which apparently is not zero. The field amplitude at this position measured using spheres of different diameters is plotted using the preliminary test data in Fig. 6c. The amplitude has a linear relationship with the sphere diameter, and the fitting line passes through the origin, confirming that the above explanation is correct.

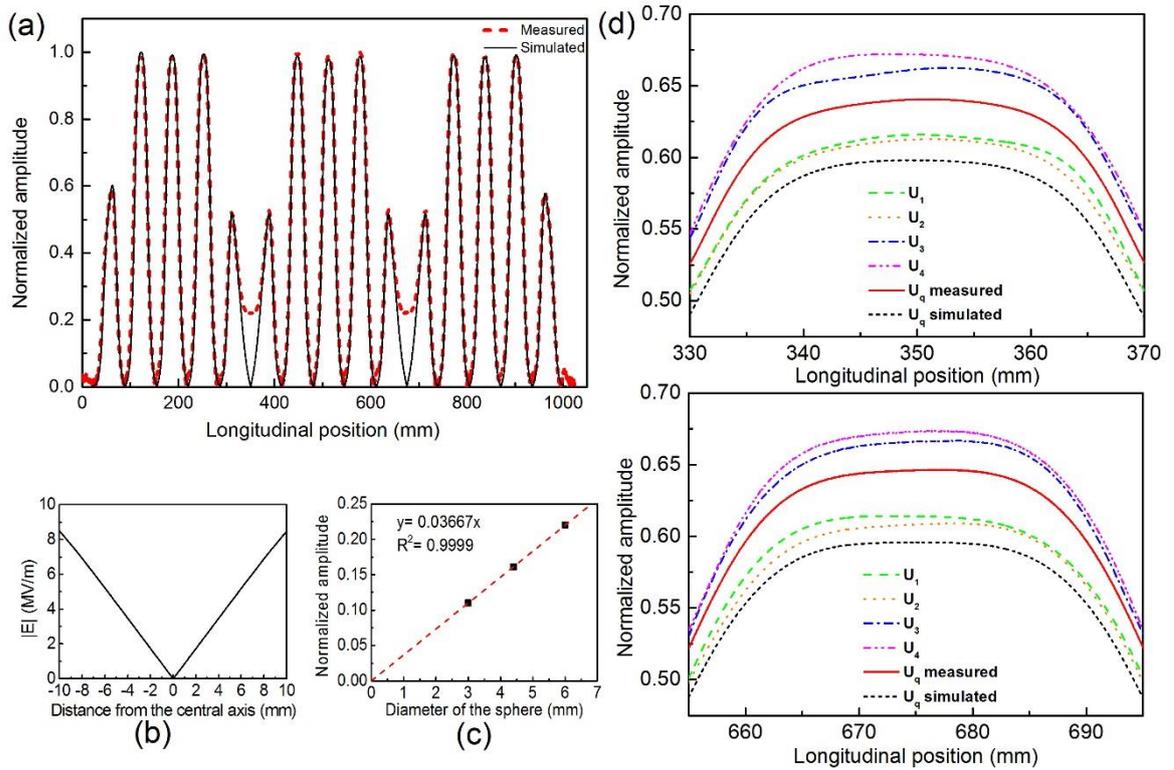

Fig. 6 RF measurement results: **a** field distribution along the central axis of the cavity (normalized by the maximum value), **b** simulated field distribution along the angle bisector of the quadrants in the middle plane of the RFQ when the stored energy of the cavity is 1 J, **c** amplitude at the center of the first RFQ section measured using spheres of different diameters, **d** measured field distribution in the four quadrants and the measured and simulated field distribution of the quadrupole field (top: first section, bottom: second section)

### 4.2.2. Distribution of the focusing field

During the measurement of the focusing field, the distance between the trajectory of the bead and the cavity axis was set to 5 mm, because the inner diameter of the drift tube is 20 mm. The field distribution along the entire cavity was measured, and the field curves corresponding to the two RFQs were extracted from the measured curves of the entire cavity. The measurement results are summarized in Fig. 6d and compared with the simulation results. The amplitudes are normalized by the maximum value over the entire trajectory of the bead, and the normalization coefficient is essentially the same as that of the accelerating field. Owing to the inhomogeneity of the nearby field, the measured amplitudes in the four quadrants are all higher than the simulated values.

For a conventional RFQ, the frequency of the operating quadrupole mode is close to that of the dipole modes, and thus accidental degeneracy between them is possible. The amplitudes of the dipole modes have to be minimized during the tuning process to produce the best approximation to a pure quadrupole mode [21]. For the spatially periodic RFQ structure, no dipole mode exists near the operating quadrupole mode because of the stem array. The amplitude of the quadrupole mode $U_q$ is defined as the average of the measured amplitudes $U_1$, $U_2$, $U_3$, and $U_4$ in the four quadrants. To evaluate the uniformity of the quadrupole field, the relative error of the quadrupole field is defined as the shape similarity error between the measured and simulated curves of $U_q$. The similarity of the curves is compared using the metrology software of the coordinate measurement machine. The relative errors of the quadrupole fields of the two RFQ sections are within ±0.4% and ±1.0%, respectively.

Distinct differences in the field amplitudes of the four quadrants can be observed in Fig. 6d. Specifically, $U_1$ and $U_2$ are relatively low, whereas $U_3$ and $U_4$ are relatively high. The differences are attributed mainly to the effects of gravity, which makes the bead sag from the desired position during the measurement [22]. Fortunately, the differences have little effect on $U_q$ because the deviations of the four quadrants' results caused by gravity cancel each other out in the calculation process.

## 5. Tuning experiment

### 5.1. Tuning algorithm

Although the measurement results show that the field errors of both the accelerating and focusing fields are within a small and acceptable range, a tuning experiment aimed at improving the field uniformity was also performed to investigate the tunability of the cavity. A tuning algorithm based on a response matrix is employed for this new structure, which is built assuming a linear relationship between each tuner setting and the field amplitude at any given longitudinal position [23]. It can be expressed by the following formula:

$$V_{di}\text{-}V_{0i} = \sum_{j=1}^{n} \frac{\partial V_i}{\partial T_j}(T_{dj} - T_{0j})$$

where $V_{di}$ is the desired field amplitude at the $i$th measurement position, and $V_{0i}$ is the actual field amplitude obtained from the measurement. $T_{dj}$ and $T_{0j}$ represent the unknown desired setting and the actual setting of the $j$th tuner, respectively. Each $\partial V_i/\partial T_j$ value is the measured derivative of the $i$th amplitude with respect to the $j$th tuner, which describes the sensitivity of the $j$th tuner to adjust the $i$th amplitude. The formula above can be rearranged in matrix form as follows:

$$V = M \cdot T$$

where $V$ and $T$ correspond to the variations of the field amplitudes and tuner settings, respectively. In addition, $M_{ij} = \partial V_i/\partial T_j$ is the response matrix. The dimensions of the response matrix are determined by the number of measurement positions and tuners. For this prototype, the measurement positions corresponding to the accelerating field and focusing field should both be included in the matrix. However, the desired values of the quadrupole field amplitudes are difficult to predict accurately owing to the systematic errors of the measurement, as mentioned in Section 4.2.1. Considering that the defined unevenness of the accelerating field is sufficient to characterize the uniformity of the overall field distribution, only the measurement positions corresponding to the accelerating field are selected, and the response matrix $M$ is defined as a $9 \times 20$ non-square matrix.

To establish the response matrix, 20 slug tuners were individually inserted 5 mm into the cavity. After each tuner was inserted, the accelerating field was measured to obtain the field variations at the measurement positions. Singular value decomposition can be used to invert the response matrix [24].

## 5.2. Tuning results

### 5.2.1. Field tuning

The normalized peak amplitudes in all the accelerating gaps after the first tuning iteration are plotted in Fig. 7a with those of the simulated and measured results before tuning. The field unevenness defined by the amplitudes in the nine gaps between the internal drift tubes is still 1.3%. However, the field unevenness is within 0.8% if the eighth gap is not considered. Furthermore, the peak amplitudes in the gaps at both ends of the cavity and the gaps between the drift tubes and the RFQs indicate that a more uniform field distribution along the entire cavity was obtained after the first tuning iteration. After the second tuning iteration, the peak amplitude in the eighth gap remained almost unchanged according to the measurement results. An examination of the response matrix revealed that the quantified tuning capability of each tuner on the eighth gap is one order of magnitude smaller than that of the tuners on the other gaps. After the first tuning iteration, the relative errors of the quadrupole fields of the two RFQ sections were within ±0.7% and ±0.8%, respectively.

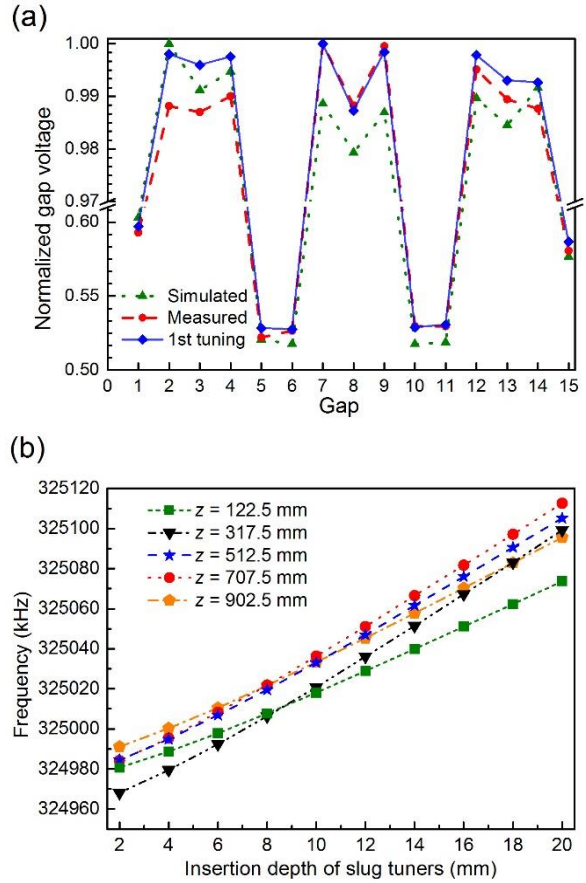

Fig. 7 **a** Peak amplitude distribution of all the accelerating gaps, **b** frequency responses of the tuners at different longitudinal positions

### 5.2.2. Frequency tuning

The frequency of the cavity was tuned after field tuning by inserting all the tuners to the same depth to maintain the field distribution. The tuners at the same longitudinal position have the same effect on the cavity frequency; therefore, only the five tuners in the first quadrant were used to obtain the frequency response curves. As shown in Fig. 7b, the cavity frequency and the insertion depth of the tuner seem to have a linear relationship when the insertion depth is greater than 6 mm. When the insertion depth is less than 6 mm, the linear relationship is invalid because the head of the tuner was machined with a fillet of 6 mm. For tuners at different longitudinal positions, the tuning coefficients of the tuners at both ends of the cavity are smaller than those of the tuners in the middle. The cavity frequency was tuned to 324.979 MHz by pulling all the tuners out by 11.02 mm. Then, the field distribution was confirmed without significant changes.

### 6. Summary

An aluminum prototype of the spatially periodic RFQ structure was designed, built, and tested to determine the accelerating and focusing field distributions, tuning performance, and mechanical structure and process technology of the cavity. Mechanically, the cavity is an overconstrained structure owing to the array of stems; thus, it requires high-precision machining and is difficult to assemble. The use of threaded connections between the drift tube components and the cavity walls mitigates the issue of overconstraint and enables manual operation of the drift tube components during the assembly process, thereby reducing the risk of assembly failure. A method of processing the drift tube components was proposed to ensure precise processing. After the cavity was assembled, the concentricity of the drift tubes measured by a laser tracker was $\phi 0.22$ mm. The rf measurement results of the prototype cavity showed good agreement with the simulation overall. The unevenness of the accelerating field was 1.3%, and the relative errors of the quadrupole fields of the two RFQ sections were within $\pm 0.4\%$ and $\pm 1.0\%$, respectively. A more uniform field distribution was obtained after a tuning experiment.

The successful fabrication and testing of the aluminum prototype provided valuable experience for the development of the spatially periodic RFQ structure. Next, the thermal performance of the cavity will be analyzed, and the structure and fabrication technology for a full-power prototype with vacuum seals and cooling channels will be studied.

## Acknowledgments

The authors would like to thank Lanzhou Ruiyuan Machinery Equipment Co. Ltd. for collaboration on the mechanical design and manufacture of the cavity. Thanks are also due to Dr. Shichun Huang for language revision. This work was supported by the NSAF Joint Foundation of China (Grant No. U1730122).

## Author contributions

Zhijun Wang and Yuan He conceived the idea of the study. The design of mechanical structure and process scheme was performed by Bin Zhang, Fengfeng Wang and Peiyan Yu. Guozhen Sun implemented the inspection of the cavity. The cold test was performed by Peiyan Yu and Chenxing Li. The remaining authors provided valuable suggestions and support for this study. The initial draft of the manuscript was written by Peiyan Yu, and all authors contributed to the writing and revisions of the manuscript.

## References

[1] I.M. Kapchinsky, V.A. Teplyakov, A linear ion accelerator with spatially uniform hard focusing. Prib. Tekh. Eksp. 2, 19 (1970).
[2] Y. He, High intensity RFQs: Review on recent developments, common problems, solutions, in: Proc. 8th Int. Particle Accelerator Conf, IPAC2017, Copenhagen, Denmark, **2017**, available at

https://accelconf.web.cern.ch/ipac2017/talks/wexa1_talk.pdf. (Accessed 6 September 2020).

[3] V.A. Teplyakov, RFQ focusing in lincas, in: Proc. 16th Linear Accelerator Conf, LINAC1992, Ottawa, Canada, 1992, pp. 21-24.

[4] Yu. Budanov, O.K. Belyaev, S.V. Ivanov et al., RFQ drift-tube proton linacs in IHEP, in: Proc. 22nd Linear Accelerator Conf, LINAC2004, Lubeck, Germany, 2004, pp. 285-287.

[5] D.A. Swenson, RF-focused drift tube linac structure, in: Proc. 17th Linear Accelerator Conf, LINAC1994, Tsukuba, Japan, 1994, pp. 804-806.

[6] D.A. Swenson, An RF focused interdigital ion accelerating structure, In: AIP Conference Proceedings 680, 1013 (2003).

[7] P.N. Ostroumov, A.A. Kolomiets, S. Sharma et al., An innovative concept for acceleration of low-energy low-charge-state heavy-ion beams. Nucl. Instrum. Methods Phys. Res. A 547, 259-269 (2005). **https://doi.org/10.1016/j.nima.2005.03.138**

[8] Z.J. Wang, Y. He, S.H. Liu et al., A new compact structure for a high intensity low-energy heavy-ion accelerator. Chin. Phys. C 37(12), **127001** (2013). **http://doi.org/10.1088/1674-1137/37/12/127001**

[9] A.A. Kolomiets, A.S. Plastun, Spatially periodic radio-frequency quadrupole focusing linac. Phys. Rev. ST Accel. Beams 18(12), **120101** (2015). **https://doi.org/10.1103/PhysRevSTAB.18.120101**

[10] G.Q. Xiao, H.S. Xu, S.C. Wang, HIAF and CiADS national research facilities: progress and prospect[J]. Nucl. Phys. Rev. 34, 275-283 (2017). **https://doi.org/10.11804/NuclPhysRev.34.03.275** (**in Chinese**)

[11] J.C. Yang, J.W. Xia, G.Q. Xiao et al., High Intensity heavy ion Accelerator Facility (HIAF) in China. Nucl. Instrum. Methods Phys. Res. B 317, 263-265 (2013). **https://doi.org/10.1016/j.nimb.2013.08.046**

[12] X.F. Niu, F. Bai, X.J. Wang et al., Cryogenic system design for HIAF iLinac. Nucl. Sci. Tech. 30, 178 (2019). **https://doi.org/10.1007/s41365-019-0700-5**

[13] U. Ratzinger, H-type linac structures, CERN Accelerator School, CERN-2005-003, 2005, pp. 351-380, available at https://cds.cern.ch/record/865926/files/p351.pdf. (Accessed 6 September 2020).

[14] Z.L. Zhang, Y. He, A.M. Shi et al., Development of the injector Ⅱ RFQ for China ADS project, in: Proc. 5th int. Particle Accelerator Conf, IPAC2014, Dresden, Germany, **2014**, pp. 3280-3282.

[15] J.D. Yuan, Y. He, B. Zhang et al., Alignment of beam position monitors in cryomodule of CADS injector Ⅱ. Nucl. Sci. Tech. 28, 75 (2017). **https://doi.org/10.1007/s41365-017-0232-9**

[16] L.C. Maier, J.C. Slater, Field strength measurements in resonant cavities. Journal of Applied Physics 23, 68-77 (1952). **https://doi.org/10.1063/1.1701980**

[17] H. Klein, Basic concepts I, CERN Accelerator School, CERN-92-03, 1992, pp. 112-118, available at https://cds.cern.ch/record/400738/files/p97.pdf. (Accessed 6 September 2020).

[18] H. Du, Y.J Yuan, Z.S. Li et al., Beam dynamics, RF measurement, and commissioning of a CW heavy ion IH-DTL. Nucl. Sci. Tech. 29, 42 (2018). **https://doi.org/10.1007/s41365-018-0373-5**

[19] Agilent Technologies, Understanding and Improving Network Analyzer Dynamic Range, Application Note 1363-1, available at http://anlage.umd.edu/Microwave%20Measurements%20for%20Personal%20Web%20Site/5980-2778EN.pdf. (Accessed 6 September 2020).

[20] H.C. Liu, J. Peng, K.Y. Gong et al., The design and construction of CSNS drift tube linac. Nucl. Instrum. Methods Phys. Res. A 911, 131-137 (2018). **https://doi.org/10.1016/j.nima.2018.10.034**

[21] C.X. Li, Y. He, F.F Wang et al., Radio frequency measurements and tuning of the China Material Irradiation Facility RFQ. Nucl. Instrum. Methods Phys. Res. A 890, 43-50 (2018). **https://doi.org/10.1016/j.nima.2018.02.047**

[22] Q. Fu, K. Zhu, Y.R. Lu et al., Detailed study of RF properties of cold models for CW window-

type RFQ. Nucl. Sci. Tech. 29, 157 (2018). **https://doi.org/10.1007/s41365-018-0489-7**

[23] T.P. Wangler, RF Linear Accelerators, John Wiley & Sons, New York, 2008.

[24] B. Koubek, A. Grudiev, M. Timmins, rf measurements and tuning of the 750 MHz radio frequency quadrupole. Phys. Rev. Accel. Beams 20(8), **080102** (2017). **https://doi.org/10.1103/PhysRevAccelBeams.20.080102**